\documentclass[onecolumn, superscriptaddress,nofootinbib,floatfix, amsfonts, aps]{revtex4}

\usepackage{amsmath}
\usepackage{bm}
\usepackage{graphicx}
\usepackage{mathrsfs}
\usepackage{footmisc}
\usepackage{multirow}
\usepackage{graphicx}
\usepackage{booktabs, ctable}
\usepackage{xcolor, color, framed}

\begin{document}

\title{Thermal Production of Charmonia in Pb-Pb Collisions at $\sqrt{s_{NN}}=5.02$ TeV}
\author{Baoyi Chen}
\affiliation{Department of Physics, Tianjin University, Tianjin 300350, China}
\affiliation{Institut f\"ur Theoretische Physik, Goethe-Universit\"at Frankfurt,
Max-von-Laue-Str. 1, D-60438 Frankfurt am Main, Germany}

\date{\today}

\begin{abstract}
This work studies the thermal production of $J/\psi$ and $\psi(2S)$ with Boltzmann transport model in the 
Quark Gluon Plasma (QGP) produced in $\sqrt{s_{NN}}=5.02$ TeV Pb-Pb collisions. $J/\psi$ nuclear modification 
factors are studied in details with the mechanisms of primordial production and the recombination of charm 
and anti-charm quarks in the thermal medium. $\psi(2S)$ binding energy is much smaller in the hot medium 
compared with the ground state, so $\psi(2S)$ with middle and low $p_T$ can 
be mainly thermally regenerated in the later stage of QGP expansions 
which enables $\psi(2S)$ inherit larger collective flows 
from the bulk medium. We quantitatively study both 
nuclear modification factors of $J/\psi$ and $\psi(2S)$ 
in different centralities and transverse momentum bins in $\sqrt{s_{NN}}=5.02$ TeV Pb-Pb collisions. 

\end{abstract}
\pacs{ }
\maketitle
\section{introduction}
Heavy flavors due to their large masses, have unique advantages in both
experimental and theoretical studies of Quantum Chromo-Dynamics (QCD).
Since $J/\psi$ was proposed as a probe of the deconfined
matter called ``Quark-gluon Plasma'' (QGP)~\cite{Matsui:1986dk}, its yield 
abnormal suppression by partons and new enhancement from the recombination
of charm and anti-charm quarks in QGP have been widely studied
in experiments~\cite{Adare:2006ns,Abelev:2012rv,Chatrchyan:2012np,
 Adam:2015jsa} and theoretical models~\cite{
Grandchamp:2001pf,xingbo:Zhao,Du:2018wsj, Zhu:2004nw,Chen:2012gg, Blaizot:2015hya,Blaizot:2018oev}.
Charmonium produced by initial hard process labelled as ``primordial production" 
at the hadronic collisions 
scatters with nucleon spectators~\cite{Gerschel:1988wn}. 
Charmonium states are usually assumed to be formed before QGP reaching local 
equilibrium, and then suffer 
inelastic
scatterings and color screening effect when charmonium move through QGP, which results 
in dissociations and
also transitions between different eigenstates ($J/\psi$,
$\psi^{\prime}$, $\chi_c$)~\cite{Satz:2005hx,Katz:2015qja,
Blaizot:2017ypk, Chen:2016mhl, Yao:2017fuc, Yao:2018nmy,Yao:2018sgn}. These eigenstates
are finally detected in experiments through the decay into dileptons.
At the LHC energies, abundant charm pairs
are produced in nuclear collisions which significantly enhances the
combination probability of $c$ and $\bar c$ to generate
new $J/\psi$s in QGP~\cite{Thews:2000rj,Andronic:2003zv,Yan:2006ve} called ``regeneration". 
As most of charm quarks are distributed in low and middle $p_T$ region, 
the regeneration process 
dominates nuclear modification factor and the collective flows of $J/\psi$ in 
low and middle $p_T$ bins~\cite{ Zhao:2011cv}. 
In high $p_T$ bin, charmonia are mainly from the 
initial hadronic collisions~\cite{Chen:2012gg}. 

More experimental data about charmonium excited state $\psi(2S)$ have 
been measured at $\sqrt{s_{NN}}=2.76$ TeV~\cite{Khachatryan:2014bva} 
and $5.02$ TeV~\cite{Sirunyan:2016znt} Pb-Pb collisions in different 
centralities and transverse momentum bins. 
$R_{AA}^{\psi(2S)}/R_{AA}^{J/\psi}$ are presented with large discrepancies 
at these two colliding energies. 
At $2.76$ TeV, $R_{AA}^{\psi(2S)}/R_{AA}^{J/\psi}$ 
becomes larger than unity in the most central collisions in $3<p_T<30$ GeV/c. 
At $5.02$ TeV, the ratio is around 
$\sim 0.5$ in a similar centrality and momentum bin. 
Both of the experimental data carry large error bars, which may prevent any 
solid conclusions. Different from $J/\psi$, $\psi(2S)$ is a loosely bound state. Its 
wavefunction is significantly modified by the hot medium which makes its 
dissociation and regeneration rates a little indistinct in the hot medium.  
With smaller binding energy, 
$\psi(2S)$ is thermally produced in the lower temperature region than $J/\psi$ and 
inherits larger collective flows from the bulk medium~\cite{Chen:2016mhl, 
Du:2015wha,  Zhao:2017yan}. 
This sequential regeneration can affect the $p_T$ dependence of the ratio 
$R_{AA}^{\psi(2S)}/R_{AA}^{J/\psi}$.  

In this work, I employ the two-component transport model to study both $J/\psi$ and 
$\psi(2S)$ production in different centralities and momentum bins at $\sqrt{s_{NN}}=5.02$ 
TeV Pb-Pb collisions. 
I updated the decay rates of excited states with a more realistic formula instead of 
a survival temperature $T_d$ above which no charmonia can survive. This 
improvement can explain well the ratio of $\psi(2S)/J/\psi$ at 5.02 TeV Pb-Pb collisions. 
In Sec.II, I introduce the details of improved Boltzmann transport model for charmonium 
evolutions and hydrodynamic equations for QGP expansion. In Sec. III, realistic 
calculations for $J/\psi$ and $\psi(2S)$ at $\sqrt{s_{NN}}=5.02$ TeV Pb-Pb 
collisions are presented and compared 
with the experimental data. A final summary is given in Sec. IV.

\section{Transport model and Hydrodynamics}

\begin{figure}[!hbt]
\centering
\includegraphics[width=0.39\textwidth]{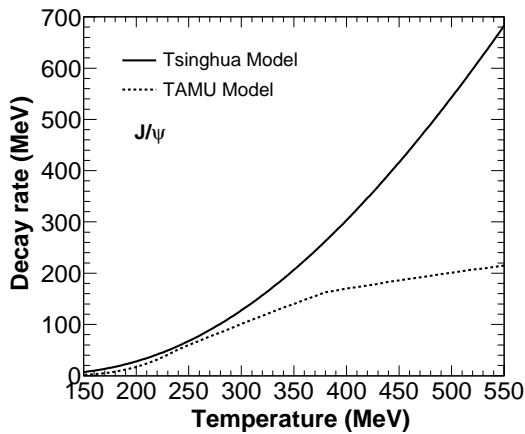}
\caption{ $J/\psi$ decay rate in the 
thermal medium as a function of temperature T. Decay rate from 
quasifree dissociation is plotted for comparison. Solid line 
is from the improved version of transport model developed by TSINGHUA 
Group~\cite{Zhu:2004nw,Chen:2012gg,Shi:2017qep}, 
and dotted line is from the calculation of TAMU 
Group~\cite{Grandchamp:2001pf,xingbo:Zhao,Du:2015wha}. 
}  
\hspace{-0.1mm}
\label{lab-decayrate}
\end{figure}

Heavy quarkonium evolutions in phase space have been well studied 
in the hot deconfined medium with 
Boltzmann transport models from SPS~\cite{Zhu:2004nw}
to the LHC~\cite{Chen:2016dke,Zhou:2014kka} in 
both p-Pb and Pb-Pb collisions. Focusing on hot medium effects, 
one can start quarkonium evolutions after their hard production. 
Three-dimensional transport equation for charmonium evolutions is simplified as, 
{\footnotesize
\begin{equation}
\label{eq-trans}
\left[\cosh(y-\eta){\frac{\partial}{\partial\tau}}+{\frac{\sinh(y-\eta)}{\tau}}{\frac{\partial}{\partial
\eta}}+{\bf v}_T\cdot\nabla_T\right]f_\Psi
=-\alpha_\Psi f_\Psi+\beta_\Psi
\end{equation}}
$f_\Psi$ is the $\Psi$ phase space density. 
y and $\eta$ are the rapidities in momentum and coordinate 
space. ${\bf v}_T={\bf p}_T/E_T={\bf p}_T/\sqrt{m_\Psi^2+p_T^2}$ 
is the transverse velocity of charmonium, 
which represents 
leakage effect in the cooling system with a finite size, i.e., charmonia with a large 
velocity tend to escape from the thermal medium instead of being dissociated.  
Primordially produced charmonia in the initial hadronic collisions suffer 
color screening effects and parton inelastic scatterings, 
both included in the decay rate $\alpha_\Psi$, 
\begin{equation}
\label{Cdfactor}
\alpha_\Psi ={1\over E_T} \int {d^3{ k}\over {(2\pi)^3E_g}}\sigma_{g\Psi}({\bf p},{\bf k},T)F_{g\Psi}({\bf p},{\bf k})f_g({\bf k},T)
\end{equation}
where $E_g$ and $f_g$ 
are gluon energy and density in the thermal medium. $F_{g\Psi}$ is the flux factor. 
In the expanding QGP, 
$u^\mu$ represents four velocity of the fluid. 
Gluon-$\Psi$ cross section in vacuum is extracted from the perturbative 
calculation with the Coulomb potential approximation. In the thermal medium, I follow Ref.\cite{Zhu:2004nw} and 
take a similar form with a reduced binding energy for charmonium,
\begin{align}
\label{sigma-gpsi}
\sigma_{g\Psi}(w)= A_0 {(w/\epsilon_\Psi-1)^{3/2}\over (w/\epsilon_\Psi)^5} 
\end{align}
with $A_0=(2^{11}\pi/27)(m_c^3\epsilon_\Psi)^{-1/2}$ and $\epsilon_\Psi$ 
to be the binding energy of $\Psi$. 
Charm quark mass is taken as the mass of D meson to fit the binding energy of charmonium 
in vacuum. 
$w=p_\Psi^\mu p_{g\mu}/m_\Psi$ 
is the gluon energy in $\Psi$ rest frame. 
In Fig.\ref{lab-decayrate}, $J/\psi$ decay rate $\alpha_\Psi$ is compared with 
the quasifree dissociation~\cite{Du:2015wha}. 
Most of $J/\psi$s can survive in the relatively low temperature region, then 
two transport 
models with these decay rates 
present similar final results in $T<300$ MeV where most of QGP and 
charmonia are located~\cite{xingbo:Zhao, Zhou:2014kka}. 
 
Heavy quark potential $V(r,T)$ can be screened partially in the thermal medium, especially 
at the large distance and high temperature supported by Lattice QCD calculations~\cite{Burnier:2014ssa}. 
Charmonium bound states may disappear sequentially in the static medium. 
The maximum survival temperature of 
a certain bound state is called ``dissociation temperature'' $T_d$, above which this 
bound state disappears. In nuclear collisions, assuming no bound states can survive at $T>T_d$ 
strongly suppresses the $\psi(2S)$ production where no 
excited states can survive inside QGP at $T>T_d^{\chi_c,\psi(2S)}\approx 1.1T_c$. 
As in the fast cooling system, charmonium states 
may also survive from the region $T>T_d$ as long as the medium cools down 
fast below $T_d$. 
In this work, I replace the approximation of $\alpha_\Psi(T>T_d)=+\infty$ employed 
in~\cite{Chen:2013wmr} 
with a large but finite value, shown in Fig.\ref{lab-decayrate}. 
The new decay rate enhances the survival probability of excited states, and 
weakly affects $J/\psi$ production because of its large $T_d$. 
$\psi(2S)$ decay rate is extracted by the geometry scale, 
$\alpha_{\psi(2S)} = \alpha_{J/\psi}\times 
\langle r\rangle^2_{\psi(2S)}/\langle r\rangle^2_{J/\psi}$, similar for $\chi_c$. Mean radius of charmonia in vacuum 
is calculated with potential model $\langle r\rangle_{J/\psi, \chi_c,\psi(2S)}
=(0.5, 0.72, 0.9)$ fm~\cite{Satz:2005hx}.

At the LHC energies, many charm pairs are also produced in Pb-Pb collisions which can 
significantly enhance the recombination of uncorrelated charm and anti-charm quark in the QGP. 
This process is included in 
Eq.(\ref{eq-trans}) with a term $\beta_\Psi$. The regeneration rate depends on both 
charm and anti-charm quark densities in the QGP, and also their recombination probability. 
At high temperature, charmonium binding energies are reduced 
significantly which suppresses the generation probability of charmonium. 
As $\psi(2S)$ is loosely 
bound, they are thermally produced in the hadronization of QGP.  
Charm quarks with color charge strongly couple with QGP and suffer energy 
loss. At relativistic heavy ion collisions, 
large quench factor and collective flows for charmed mesons have been observed
~\cite{Abelev:2006db,Adare:2006nq, Abelev:2013lca}. 
Therefore, one can approximately take kinetically thermalized phase space distribution for charm 
quarks at $\tau\ge \tau_0$ where $\tau_0$ is the time scale of QGP local equilibrium~\cite{Zhao:2017yhj}. 
As heavy quarks are barely produced from the thermal medium due to its large mass, total 
number of charm pairs is conserved with spatial 
diffusions inside QGP~\cite{Zhao:2018jlw}. 
The spatial density is controlled by the conservation equation. 
\begin{align}
\label{eq-cflow}
\partial_\mu(\rho_cu^\mu)=0
\end{align} 
The initial charm density at $\tau_0$ is obtained by nuclear geometry, 
\begin{equation}
\label{dnsty}
\rho_c({\bf x}_T, \eta,\tau_0) = {{d\sigma_{pp}^{c{\bar c}}} \over d\eta}{ T_A({\bf x}_T)T_B({\bf x}_T-{\bf b})\cosh(\eta)\over \tau_0}
\end{equation}
where $T_A$ and $T_B$ are thickness functions of two colliding nuclei, with the definition 
of $T_{A(B)}({\bf x}_T)=\int_{-\infty}^{\infty} dz\rho_{A(B)}({\bf x}_T,z)$. 
$\rho_{A(B)}({\bf x}_T,z)$ is taken as Woods-Saxon nuclear density.
The rapidity distribution of charm pairs in $\sqrt{s_{NN}}=5.02$ TeV pp 
collisions 
is obtained by the interpolation of the experimental 
data at 2.76 TeV and 7 TeV, 
$d\sigma_{pp}^{c\bar c}/dy=0.86$ mb 
in the central rapidity $|y|<0.9$ 
and $0.56$ mb in the forward rapidity $2.5<|y|<4$~\cite{Aaij:2016jht}. 

The momentum distribution of charmonium primordial production in AA collisions 
is scaled from the distribution in pp collisions. 
The parametrization of charmonium 
initial distribution at 5.02 TeV is similar to the form at 2.76 TeV, 
\begin{align}
\label{eq-pp1S}
&{d^2\sigma_{\mathrm{pp}}^{J/\psi}\over dy2\pi p_Tdp_T} = f_{J/\psi}^{\mathrm{Norm}}(p_T|y)
\cdot {d\sigma_{\mathrm{pp}}^{J/\psi}\over dy} \\
\label{eq-norm1S}
&f_{J/\psi}^{\mathrm{Norm}}(p_T|y)
={(n-1)\over {\pi(n-2)\langle p_T^2\rangle_{pp}}}
[1+{p_T^2\over {(n-2)\langle p_T^2\rangle_{pp}}}]^{-n}
\end{align}  
Charmonium rapidity differential cross section at 5.02 TeV is 
$d\sigma_{pp}^{J/\psi}/dy=5.0\ \mu b$ in central rapidity $|y|<1$ 
and $3.25\ \mu b$ in the forward rapidity $2.5<|y|<4$, through the interpolation 
between the experimental data of 2.76 TeV~\cite{Abelev:2012kr} 
and 7 TeV~\cite{Abelev:2012gx} pp collisions. 
$f_{J/\psi}^{\mathrm{Norm}}(p_T|y)$ 
is the normalized transverse momentum distribution of charmonium with rapidity $y$. 
The mean transverse momentum 
square $\langle p_T^2\rangle$ and the parameter $n$ is 
extracted to be $\langle p_T^2\rangle_{pp}|_{y=0}=12.5\ \mathrm{(GeV/c)^2}$ 
and $n=3.2$. For charmonium momentum distribution in other rapidities, 
lack of more constraints, $\langle p_T^2\rangle_{pp}(y)$ is determined by the relation, 
\begin{align}
\langle p_T^2\rangle_{pp}^{J/\psi}(y) = \langle p_T^2\rangle_{pp}^{J/\psi}|_{y=0}\times 
[1-({y\over y_{\mathrm{max}}})^2]
\end{align}
where $y_{\mathrm{max}}=\cosh^{-1}(\sqrt{s_{NN}}/(2E_T))$ is the maximum rapidity of 
charmonium in pp collisions with zero transverse momentum. 
As the masses of charmonium excited states ($\chi_c,\psi(2S)$) are close to $J/\psi$, 
their initial momentum distributions 
are approximated to be the same with Eq.(\ref{eq-pp1S}-\ref{eq-norm1S}). 

In nuclear collisions, charmonium initial distribution is also 
modified by the shadowing effects in the nucleus~\cite{Vogt:2004dh,Eskola:1998df}. 
I employ the EPS09 NLO model~\cite{Eskola:2009uj} 
to generate the modification factors  
for primordially produced charmonium at $\sqrt{s_{NN}}=5.02$ TeV Pb-Pb collisions. 
This suppression factor is roughly $\sim 0.8$ depending on the impact parameter. 
For the regeneration, shadowing effect reduces the number of charm pairs by around 
20\%, and suppress the regeneration by a factor of $\sim 0.8^2$. 

QGP expansion as a background for charmonium evolutions is simulated  
with the (2+1) 
dimensional ideal hydrodynamic equations in the transverse plane, 
with the assumption of Bjorken expansion in longitudinal direction. 
\begin{align}
\partial_\mu T^{\mu\nu}=0
\end{align}
$T^{\mu\nu}=(e+p)u^\mu u^\nu -g^{\mu\nu}p$ is the energy-momentum tensor. 
$e$ and $p$ are the energy density and the pressure. $u^\mu$ is the four 
velocity of QGP fluids, which can affect charm spatial diffusions through 
Eq.(\ref{eq-cflow}) and the charmonium regeneration. It also determines the collective flows 
of light hadrons, charmed mesons and the regenerated charmonia. 
The deconfined matter is 
treated as an ideal gas of massless gluons, $u$ 
and $d$ quarks, and strange quark with mass 
$m_s=150$ MeV~\cite{Sollfrank:1996hd}. Hadron gas is an 
ideal gas of all known hadrons and 
resonances with mass up to 2 GeV~\cite{Patrignani:2016xqp}. 
Two phases are connected with 
first-order phase transition and the critical temperature is $T_c=170$ MeV. 
The initial maximum temperature of QGP is 
extracted to be $T_0({\bf x}_T=0, \tau_0)= 510$ MeV in the central rapidity $|y|<2.4$ 
and $450$ MeV in the forward rapidity $2.5<|y|<4$. Here $\tau_0=0.6$ 
fm/c is the time scale of QGP reaching local equilibrium~\cite{Zhao:2017yhj}. 
The lifetime of QGP is $\sim 10$ fm/c in the most central Pb-Pb collisions 
at $\sqrt{s_{NN}}=5.02$ TeV.

\section{Numerical results and analysis}

%\subsection{$J/\psi$ Nuclear Modification Factors with Transverse Momentum $p_T$ and 
%Rapidity $y$} 

\begin{figure*}[!t]
\centering
\includegraphics[height=6.0cm]{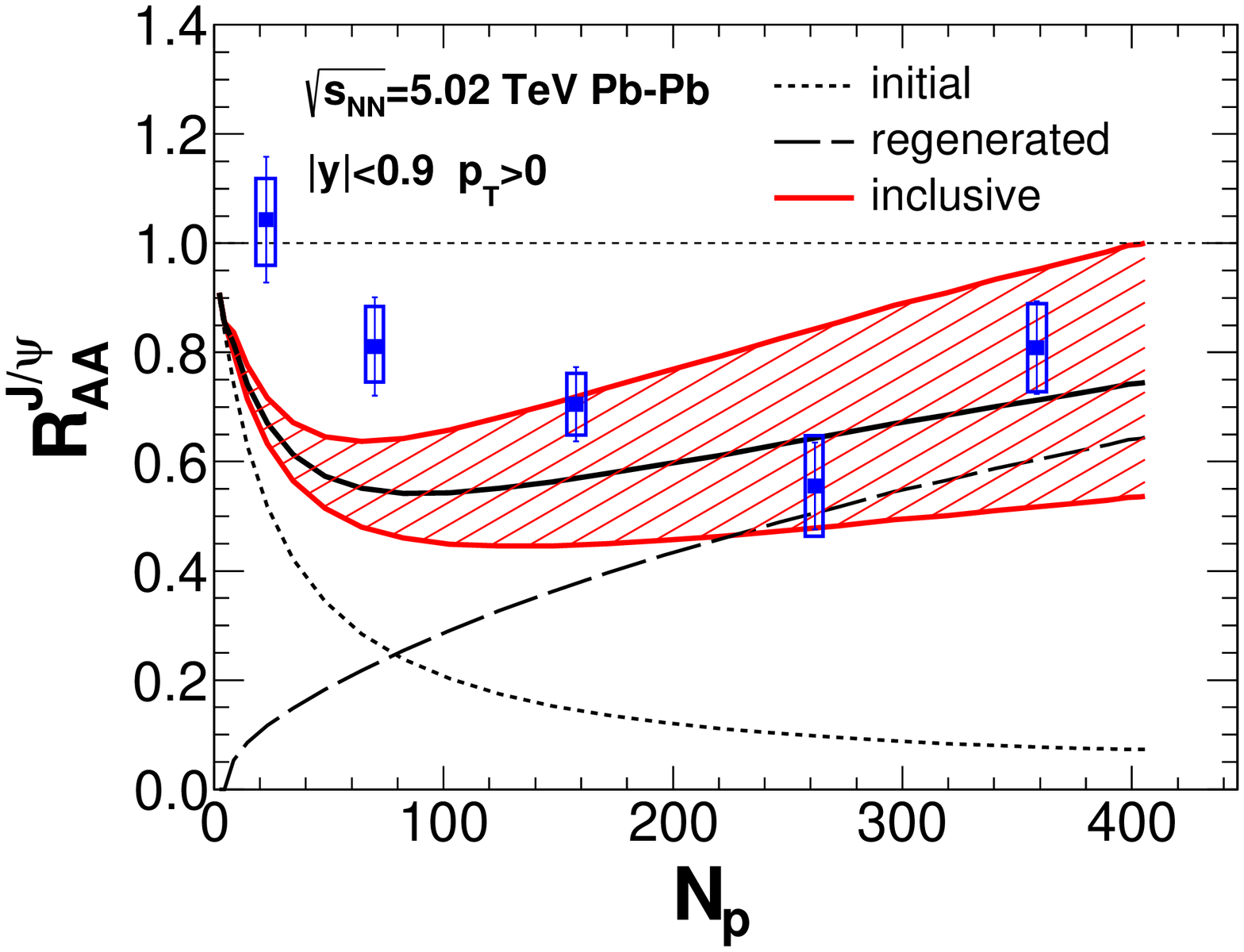}
\hbox{ } \hspace{0.3cm} \hbox{ }
\includegraphics[height=6.0cm]{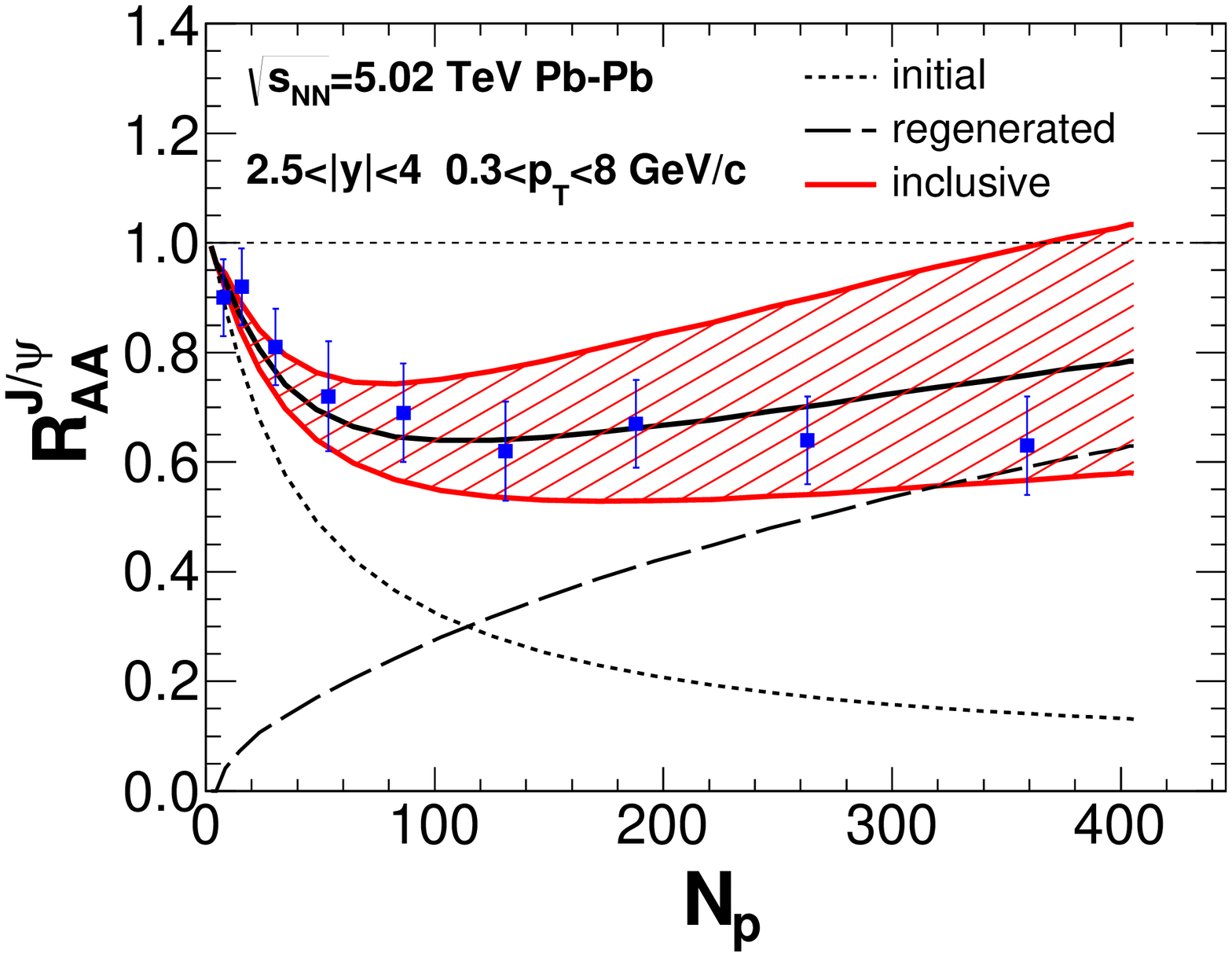}
%\includegraphics[width=0.4\textwidth]{rR-U11Tc-Gauss.eps}
%\hbox{ } \hspace{0.3cm} \hbox{ }
%\includegraphics[width=0.4\textwidth]{rR-U20Tc-Gauss.eps}
\caption{
(Left panel) inclusive nuclear modification factor $R_{AA}$ 
of $J/\psi$ as a function of the number of participants $N_p$ in central rapidity in 
Pb-Pb collisions at $\sqrt{s_{NN}}=5.02$ TeV. The dotted and dashed lines are the primordial 
production and regeneration respectively. Black solid line in the middle of the band is 
for $J/\psi$ inclusive $R_{AA}$. 
Color band is for the uncertainties of the inputs, with $d\sigma_{pp}^{c\bar c}/dy$ changed by $\pm 20\%$. 
Experimental data is from ALICE Collaboration~\cite{JimenezBustamante:2017wxc}. 
(Right panel) inclusive $R_{AA}$ in forward rapidity. 
The lines and the band 
are similar to the left panel. 
Experimental data is from ALICE Collaboration~\cite{Adam:2016rdg}. 
}
\hspace{-0.1mm}
\label{labRAA-Np-both}
\end{figure*}

With the transport model for charmonium evolutions and hydrodynamic 
equations for QGP collective expansion, one can obtain the realistic nuclear 
modification factors of charmonia in the heavy ion collisions. 
In the left panel of Fig.\ref{labRAA-Np-both}, primordially produced 
charmonia suffer dissociations from peripheral to central Pb-Pb collisions, plotted 
with the dotted line. The regeneration from $c+\bar c 
\rightarrow J/\psi +g$ is plotted with dashed line which is proportional to the number 
of charm pairs in QGP, and dominates $J/\psi$ 
total production in the central collisions. The experimental data in left panel of 
Fig.\ref{labRAA-Np-both} 
is inclusive production which includes the non-prompt part 
from B-hadron decays. It contributes around 10\% to the final inclusive yields. 
Detailed momentum dependence of 
non-prompt fraction in $J/\psi$ inclusive production is fitted as  
$f_B=N_{pp}^{B\rightarrow J/\psi}/(N_{pp}^{\mathrm{prompt}}+N_{pp}^{B\rightarrow J/\psi})
=0.04+0.023p_T/(GeV/c)$~\cite{Chen:2013wmr}, with weak 
dependence on rapidity and the colliding energies $\sqrt{s_{NN}}$. 
In nuclear collisions, bottom quarks suffer strong energy loss in the thermal medium. B hadrons and non-prompt 
charmonia are shifted from high $p_T$ to relatively low $p_T$. 
This hot medium modification on 
non-prompt charmonium (or bottom quark) momentum distribution 
is characterized with a quench factor $R_Q$. 
In high $p_T$, quench factor $R_Q$ 
is smaller than unity, extracted 
to be $0.4$ from non-prompt $J/\psi$ $R_{AA}$~\cite{Chen:2013wmr}. 
This value is also employed in the entire $p_T$ region. 
With both prompt and non-prompt charmonia, one can obtain charmonium 
inclusive nuclear modification factors in Fig.\ref{labRAA-Np-both}. 
Considering large uncertainties of $d\sigma_{pp}^{c\bar c}/dy$ in the transport model, I perform two 
calculations for $R_{AA}$ with the change of $d\sigma_{pp}^{c\bar c}/dy$ by $\pm 20\%$, 
see the color band 
in Fig.\ref{labRAA-Np-both}. 
In most central collisions, primordial production is strongly 
suppressed and regeneration dominates the total yield. 

%\begin{figure}[!hbt]
%\centering
%\includegraphics[width=0.4\textwidth]{fig2}
%\caption{(Color online) 
%The inclusive nuclear modification factor $R_{AA}$ 
%of $J/\psi$ as a function of the number of participants $N_p$ in central rapidity in 
%Pb-Pb collisions at $\sqrt{s_{NN}}=5.02$ TeV. The dotted and dashed lines are the primordial 
%production and regeneration respectively. Black solid line in the middle of the band is 
%for $J/\psi$ inclusive $R_{AA}$. 
%Color band is for the uncertainties of the inputs, with $d\sigma_{pp}^{c\bar c}/dy$ changed by $\pm 20\%$. 
%Experimental data is from ALICE Collaboration~\cite{JimenezBustamante:2017wxc}. 
%Right panel: inclusive $R_{AA}$ in forward rapidity. 
%The lines and the band 
%are similar to the left panel. 
%Experimental data is from ALICE Collaboration~\cite{Adam:2016rdg}. 
%}  
%\hspace{-0.1mm}
%\label{labRAA-Np-centY}
%\end{figure}

%%%%%%%%%%%%%%%%%%%%%%%%%
\begin{figure*}[!t]
\centering
\includegraphics[height=6.0cm]{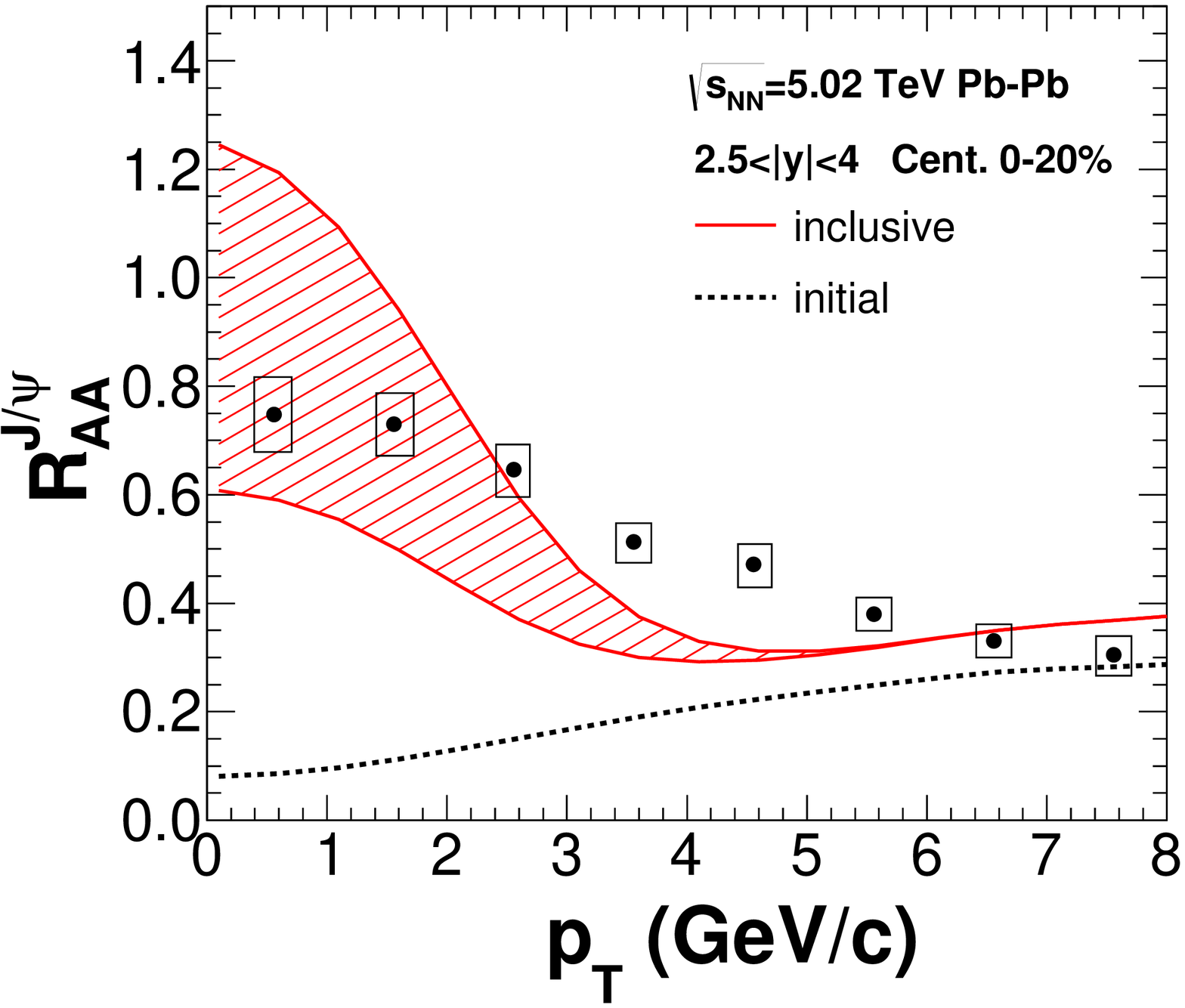}
\hbox{ } \hspace{0.3cm} \hbox{ }
\includegraphics[height=6.0cm]{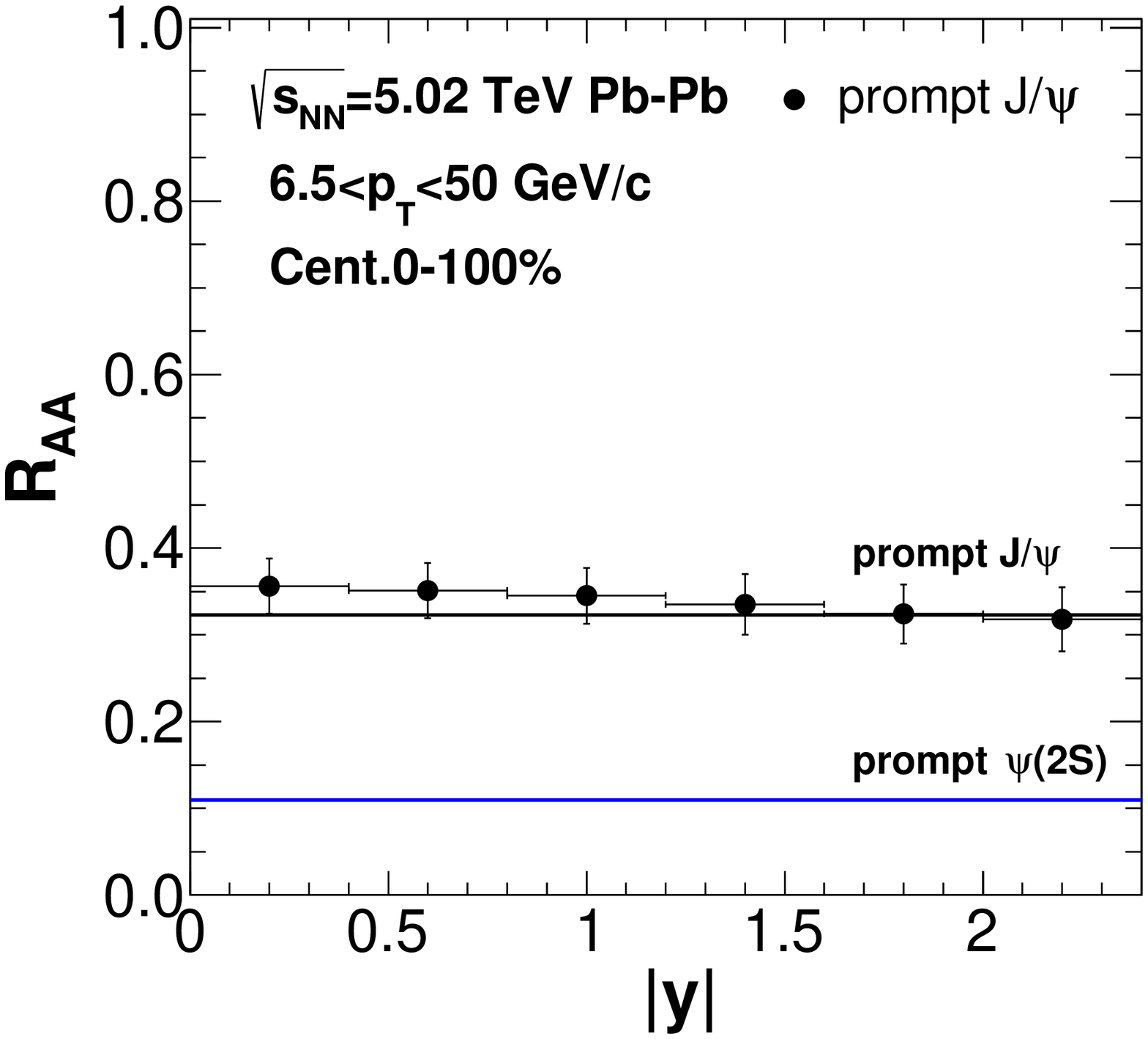}
%\includegraphics[width=0.4\textwidth]{rR-U11Tc-Gauss.eps}
%\hbox{ } \hspace{0.3cm} \hbox{ }
%\includegraphics[width=0.4\textwidth]{rR-U20Tc-Gauss.eps}
\caption{(Color online) 
(Left panel) $J/\psi$ nuclear modification factor $R_{AA}$ 
as a function of transverse momentum $p_T$. 
Dotted line is for initial production, solid line is for inclusive production 
consisting of initial production, regeneration and B hadron decay. 
Color band is due to the 
uncertainties of $d\sigma_{pp}^{c\bar c}/dy$ (see Fig.\ref{labRAA-Np-both}).  
Difference between solid line and dotted 
line is mainly due to the contribution of regeneration at low $p_T$ and B hadron decay 
at high $p_T$ respectively. 
Experimental data is from ALICE Collaboration~\cite{Paul:2016qee}. 
(Right panel) $J/\psi$ prompt $R_{AA}$ as a function of rapidity in the centrality 0-100\%. 
$\psi(2S)$ prompt $R_{AA}(y)$ is also predicted. The experimental data are from \cite{Sirunyan:2017isk}.   
}
\hspace{-0.1mm}
\label{labRAA-pT-rap}
\end{figure*}

In the forward rapidity, both initial conditions of hydrodynamic equations 
and the transport model are updated properly compared with the central rapidity 
collisions. In the right panel of Fig.\ref{labRAA-Np-both}, 
$J/\psi$ nuclear modification factor 
from the primordial production (dotted line), regeneration (dashed line), and the 
inclusive production (color band) are plotted separately. The flat tendency 
of experimental data with $N_p$ is due to the combined effects of the  
decrease of primordial production and the increase of regeneration in final $J/\psi$ 
yield. The experimental data in the right panel of Fig.\ref{labRAA-Np-both} 
is at $0.3<p_T<8$ GeV/c. It can exclude the 
contribution of coherent photoproduction which are distributed below 0.3 GeV/c. 
The additional contribution from 
charmonium 
coherent photoproduction can make total 
$R_{AA}$ larger than unit in ultra-peripheral collisions~\cite{Shi:2017qep}.

In order to show the contributions of primordial production and regeneration, 
the $p_T$-differential $R_{AA}$ is also plotted in left panel of Fig.\ref{labRAA-pT-rap}. 
Significant 
enhancement of $R_{AA}$ in the low $p_T$ region is caused by the regeneration. 
Large suppression 
in high $p_T$ region is due to the color screening and parton inelastic collisions. 
Dotted line is for the initial production, it increases slightly with $p_T$ due to 
the leakage effect. 
Both theoretical results of $R_{AA}^{J/\psi}$ in 
central and forward rapidities can explain experimental data well.  
Note that even the charm pair cross section $d\sigma_{pp}^{c\bar c}/dy$ 
in central rapidity is larger than the value in 
forward rapidity, $R_{AA}$ is similar to each other in two rapidities. Because in central rapidity with 
hotter medium, QGP strong expansion ``blows" charm quarks to a larger volume, which suppresses the 
charm quark spatial density and the charmonium regeneration. Meanwhile, the elliptic flows of regenerated 
charmonia become larger in the central rapidity. These will be discussed in details below.   
In the right panel of Fig.\ref{labRAA-pT-rap}, $J/\psi$ and $\psi(2S)$ prompt $R_{AA}$ 
as a function of rapidity is also presented. 

%\subsection{$\psi(2S)$ Thermal Production and the Ratio $\psi(2S)/J/\psi$}
%%%%%%%%%%%%%%%%%%%%%%%%%%%
\begin{figure*}[t]
\centering
\includegraphics[height=6.0cm]{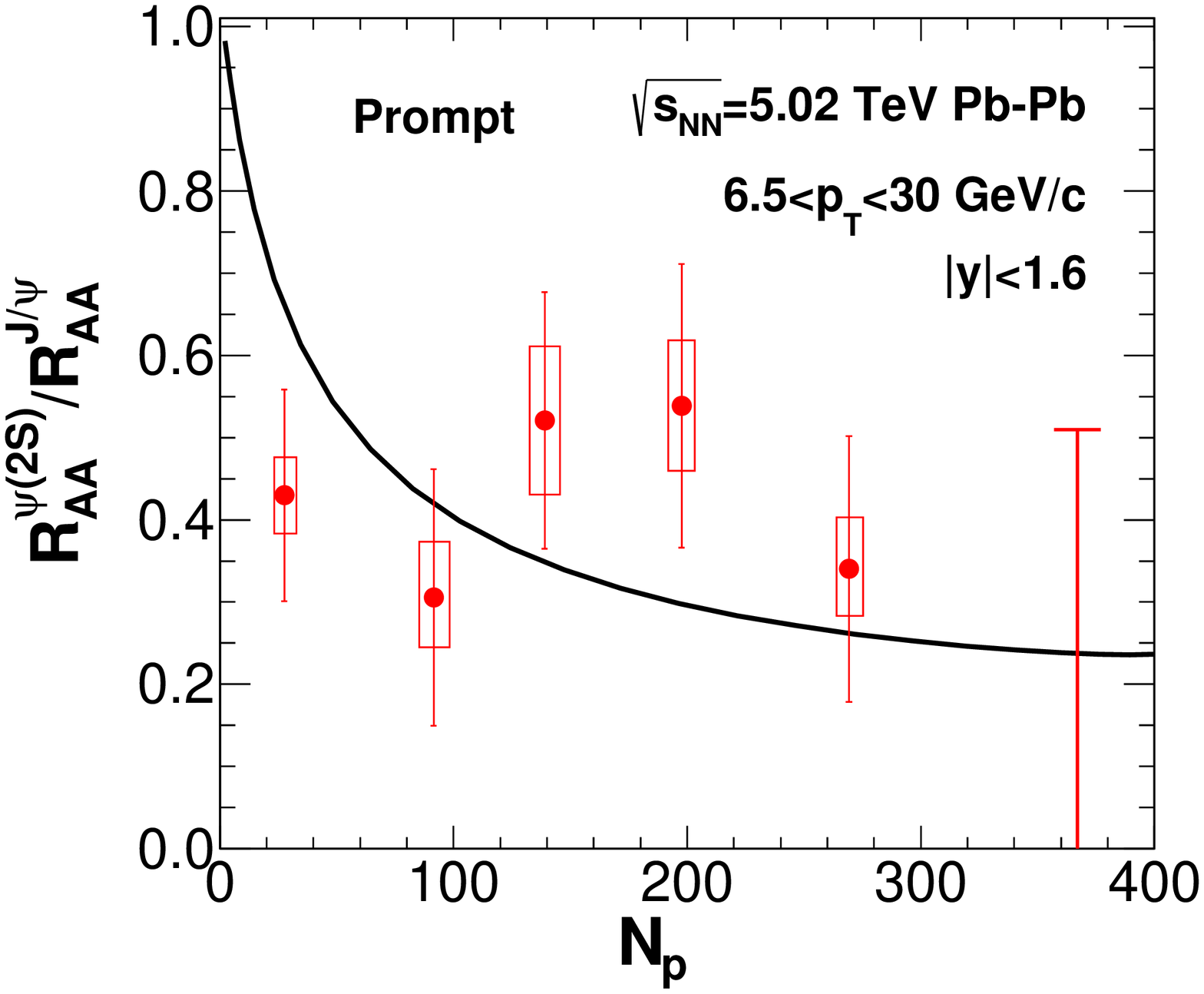}
\hbox{ } \hspace{0.3cm} \hbox{ }
\includegraphics[height=6.0cm]{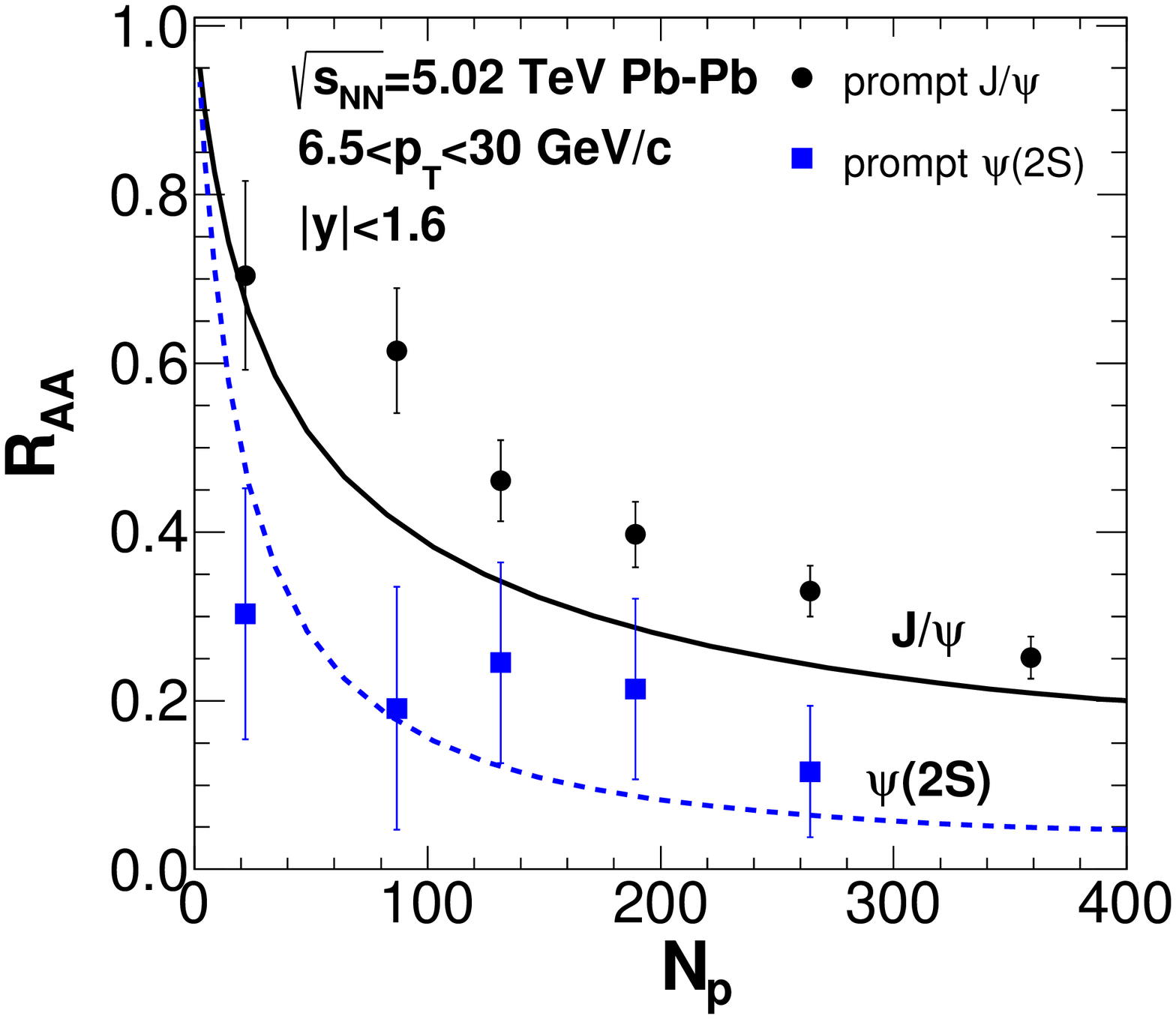}
%\includegraphics[width=0.4\textwidth]{rR-U11Tc-Gauss.eps}
%\hbox{ } \hspace{0.3cm} \hbox{ }
%\includegraphics[width=0.4\textwidth]{rR-U20Tc-Gauss.eps}
\caption{(Color online) 
(Left panel) the ratio of $J/\psi$ and $\psi(2S)$ prompt nuclear modification factors 
in central rapidity region with a momentum cut $6.5<p_T<30$ GeV/c in $\sqrt{s_{NN}}=5.02$ TeV 
Pb-Pb collisions. Experimental data are from CMS Collaboration~\cite{Sirunyan:2016znt}.
(Right panel) $J/\psi$ and $\psi(2S)$ prompt nuclear 
modification factor as a function of $N_p$  
in central rapidity region with a momentum cut $6.5<p_T<30$ GeV/c in $\sqrt{s_{NN}}=5.02$ TeV 
Pb-Pb collisions. The experimental data are from \cite{Sirunyan:2017isk}.
}
\hspace{-0.1mm}
\label{labRAA-DR-PT6-both}
\end{figure*}

%%%%%%%%%%%%%%%%%%%%%%%%%%%%%%%%%%%%
\begin{figure*}[thb]
\centering
\includegraphics[height=6.0cm]{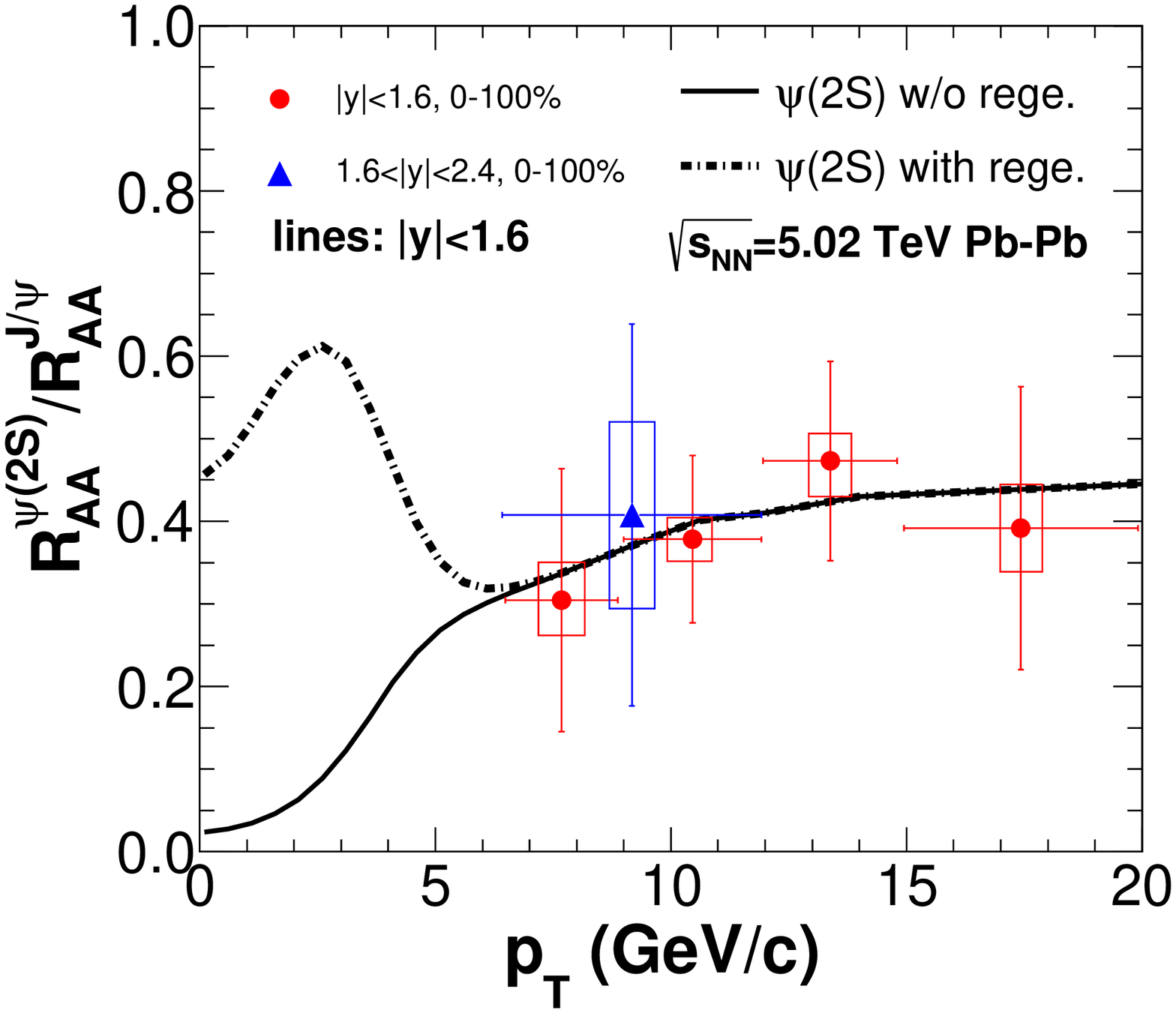}
\hbox{ } \hspace{0.3cm} \hbox{ }
\includegraphics[height=6.0cm]{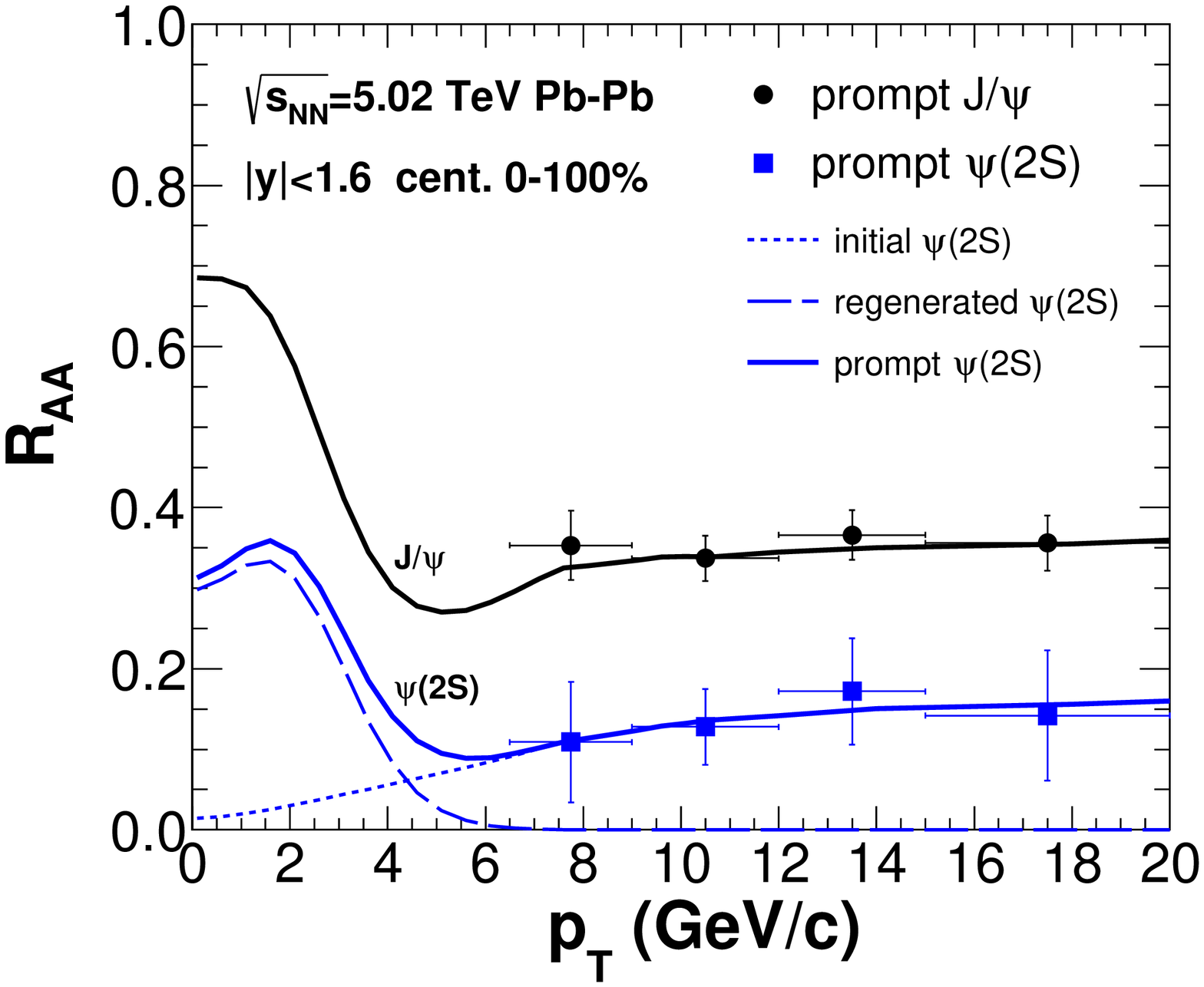}
%\includegraphics[width=0.4\textwidth]{rR-U11Tc-Gauss.eps}
%\hbox{ } \hspace{0.3cm} \hbox{ }
%\includegraphics[width=0.4\textwidth]{rR-U20Tc-Gauss.eps}
\caption{(Color online) 
(Left panel) Ratio of $J/\psi$ and $\psi(2S)$ prompt nuclear modification factors as 
a function of the transverse momentum $p_T$ in the minimum bias 
(corresponding to the impact parameter b=8.4 fm) at $\sqrt{s_{NN}}=5.02$ TeV Pb-Pb 
collisions. Dotted-dashed and solid lines are with and without $\psi(2S)$ regeneration, 
respectively. 
Experimental data are from CMS Collaboration~\cite{Sirunyan:2016znt, Sirunyan:2017isk}. 
(Right panel) $J/\psi$ and $\psi(2S)$ nuclear modification factor as a function of transverse momentum. 
Black dotted, dashed and solid lines are for initial, regenerated and prompt $\psi(2S)$ 
respectively.  
Prompt $R_{AA}^{J/\psi}$ is also plotted for comparison. 
The experimental data are from \cite{Sirunyan:2016znt, Sirunyan:2017isk}.
}
\hspace{-0.1mm}
\label{labRAA-DR-pt-var}
\end{figure*}

Situation becomes a little complicated for $\psi(2S)$ 
in the hot medium because of its dissociation rate 
compared with the tightly 
bound $J/\psi$. $\psi(2S)$ decay 
rate here is extracted from $J/\psi$' by the geometry scale in Section II. 
In Fig.\ref{labRAA-DR-PT6-both}, charmonia with high $p_T$ 
are mainly from the primordial production. From peripheral to central collisions, 
$\psi(2S)$ suffers stronger dissociation and the 
ratio of $R_{AA}^{\psi(2S)}/R_{AA}^{J/\psi}$ 
decreases with $N_p$. In the most central collisions at LHC, 
the ratio of charmonium 
nuclear modification factors is proportional to their decay rates. 
In peripheral collisions, charmonium path length in hot medium becomes smaller. With weak 
suppression, $J/\psi$ and $\psi(2S)$ nuclear modification factors approach unity, which 
makes $R_{AA}^{\psi(2S)}/R_{AA}^{J/\psi}\rightarrow 1$ at $N_p\rightarrow 2$, 
see left panel of Fig.\ref{labRAA-DR-PT6-both}. 
Individual $R_{AA}^{J/\psi}$ in high $p_T$ bin is also 
plotted in the right panel of Fig.\ref{labRAA-DR-PT6-both}. 
The decreasing tendency of $J/\psi$ and $\psi(2S)$ $R_{AA}$ with $N_p$ are explained well.

Transverse momentum dependence of $R_{AA}^{\psi(2S)}/R_{AA}^{J/\psi}$ is also studied based on the 
sequential regeneration mechanism. In 
Fig.\ref{labRAA-DR-pt-var}, 
at $p_T\rightarrow 0$, there will be significant regeneration 
for $J/\psi$. For loosely bound $\psi(2S)$, 
they can only be thermally produced in the later stage of QGP expansion compared 
with $J/\psi$, which makes regenerated $\psi(2S)$ inherit larger velocity and collective 
flows from the bulk medium based on the fact of strong couplings between charm quarks and 
the deconfined medium. Therefore, the regenerated $\psi(2S)$ are distributed in the larger 
$p_T$ (dashed line in righ panel of Fig.\ref{labRAA-DR-pt-var}) compared with the 
regenerated $J/\psi$. 
Due to the different $p_T$ distributions of thermally produced $J/\psi$ and $\psi(2S)$, 
the shapes of 
$J/\psi$ and $\psi(2S)$ $R_{AA}(p_T)$s 
(black and blue solid lines in right panel of Fig.\ref{labRAA-DR-pt-var}) 
are different.  
There is a ``peak'' in the ratio $R_{AA}^{\psi(2S)}/R_{AA}^{J/\psi}$ in the left 
panel of Fig.\ref{labRAA-DR-pt-var} due to the sequential regeneration of $\psi(2S)$. 
If without $\psi(2S)$ regeneration, the ratio will decreases to zero at $p_T\rightarrow 0$, see 
the solid line in the left panel of Fig.\ref{labRAA-DR-pt-var}.

\begin{figure}[!hbt]
\centering
\includegraphics[width=0.4\textwidth]{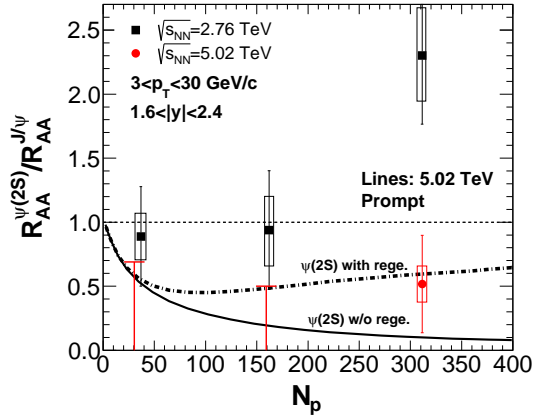}
\caption{(Color online) Ratio of $J/\psi$ and $\psi(2S)$ prompt nuclear modification factors as a 
function of $N_p$ in rapidity $1.6<|y|<2.4$ with a momentum cut $3<p_T<30$ GeV/c. 
Dotted-dashed line is the scenario for $J/\psi$ and $\psi(2S)$ with both primordial production and 
regeneration, solid line is the scenario without regeneration for $\psi(2S)$ 
($J/\psi$ regeneration is included in both lines). Experimental data at $\sqrt{s_{NN}}=2.76$ 
TeV and 5.02 TeV are from CMS Collaboration~\cite{Sirunyan:2016znt}. 
}
\hspace{-0.1mm}
\label{labDR-Np-PT3}
\end{figure}

In the low and middle $p_T$ region, regeneration becomes important 
for $J/\psi$ and $\psi(2S)$ and enhances their $R_{AA}$ in semi-central and central 
collisions. In order to show the roles of $\psi(2S)$ regeneration on $R_{AA}^{\psi(2S)}/R_{AA}^{J/\psi}$, we present two calculations with and without $\psi(2S)$ regeneration 
respectively in Fig.\ref{labDR-Np-PT3}. 
In Fig.\ref{labDR-Np-PT3}, 
neglecting the regeneration for $\psi(2S)$ (solid line), 
$R_{AA}^{\psi(2S)}/R_{AA}^{J/\psi}$ keeps 
dropping down with $N_p$ due to the stronger QGP suppression on charmonium excited 
states. The supplement of regeneration increases $\psi(2S)$ production especially 
in the central collisions and enhances the value of $R_{AA}^{\psi(2S)}/R_{AA}^{J/\psi}$, 
see the dotted-dashed line. Note that  
in the work of Ref.\cite{Chen:2013wmr}, predictions about prompt
$R_{AA}^{\psi(2S)}/R_{AA}^{J/\psi}$ 
at 2.76 TeV Pb-Pb collisions have been made. Its 
value is predicted to be around $\sim 0.15$ in all $p_T$ bins. The binding energy and 
the regeneration rate for $\psi(2S)$ 
in Ref.\cite{Chen:2013wmr} is smaller which suppress
$\psi(2S)$ production. I extend previous calculations from 2.76 TeV~\cite{Chen:2013wmr} 
to 5.02 TeV, both of them are consistent with the experimental data at 5.02 TeV. 
The difference between  
solid and dotted-dashed 
lines in Fig.\ref{labDR-Np-PT3} are due to the $\psi(2S)$ regeneration component. 

%%%%%%%%%%%%%%%%%%%%%%%%%

%\subsection{Elliptic Flows of Sequentially Produced $J/\psi$ and $\psi(2S)$}

\begin{figure*}[!t]
\centering
\includegraphics[height=6.0cm]{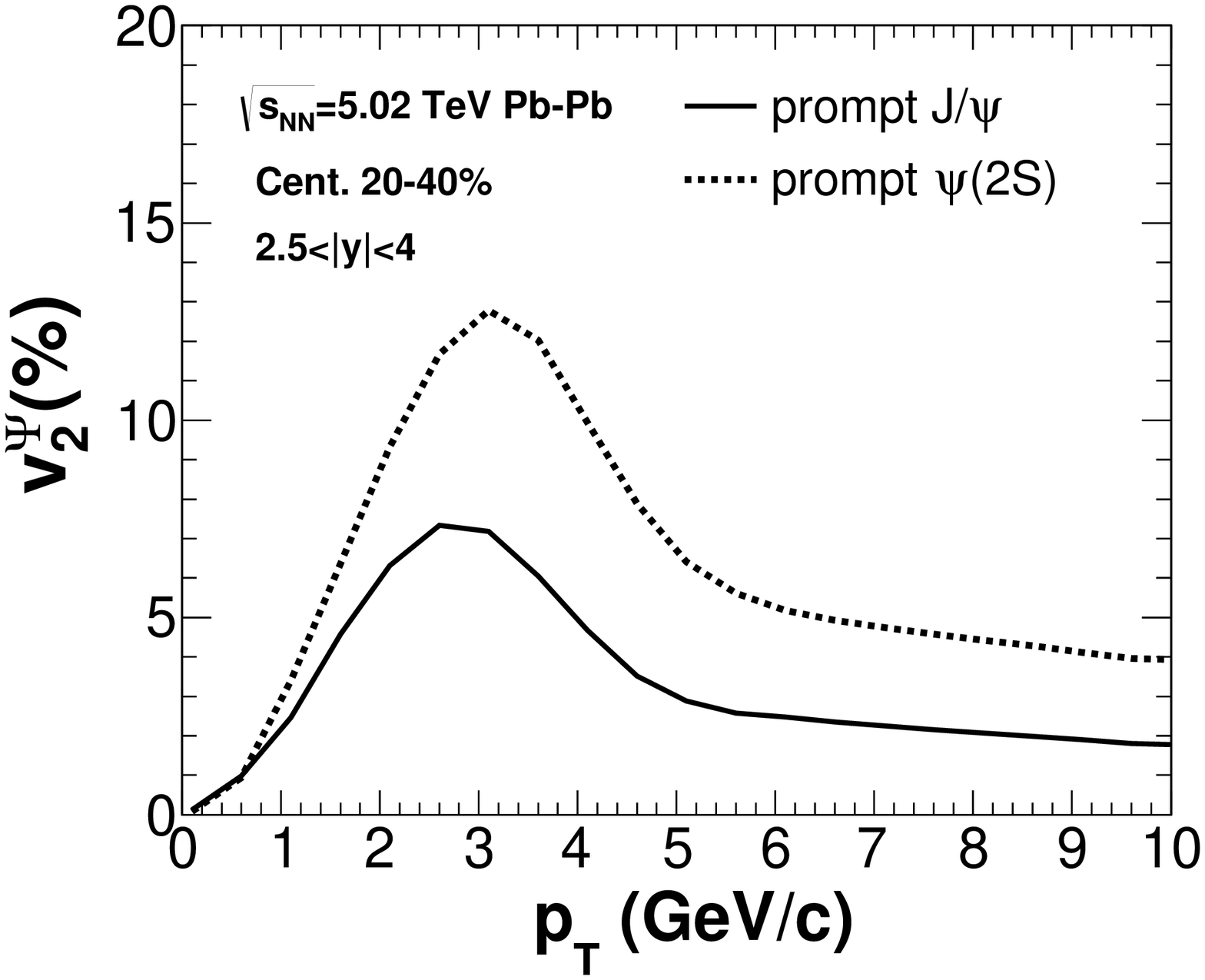}
\hbox{ } \hspace{0.3cm} \hbox{ }
\includegraphics[height=6.0cm]{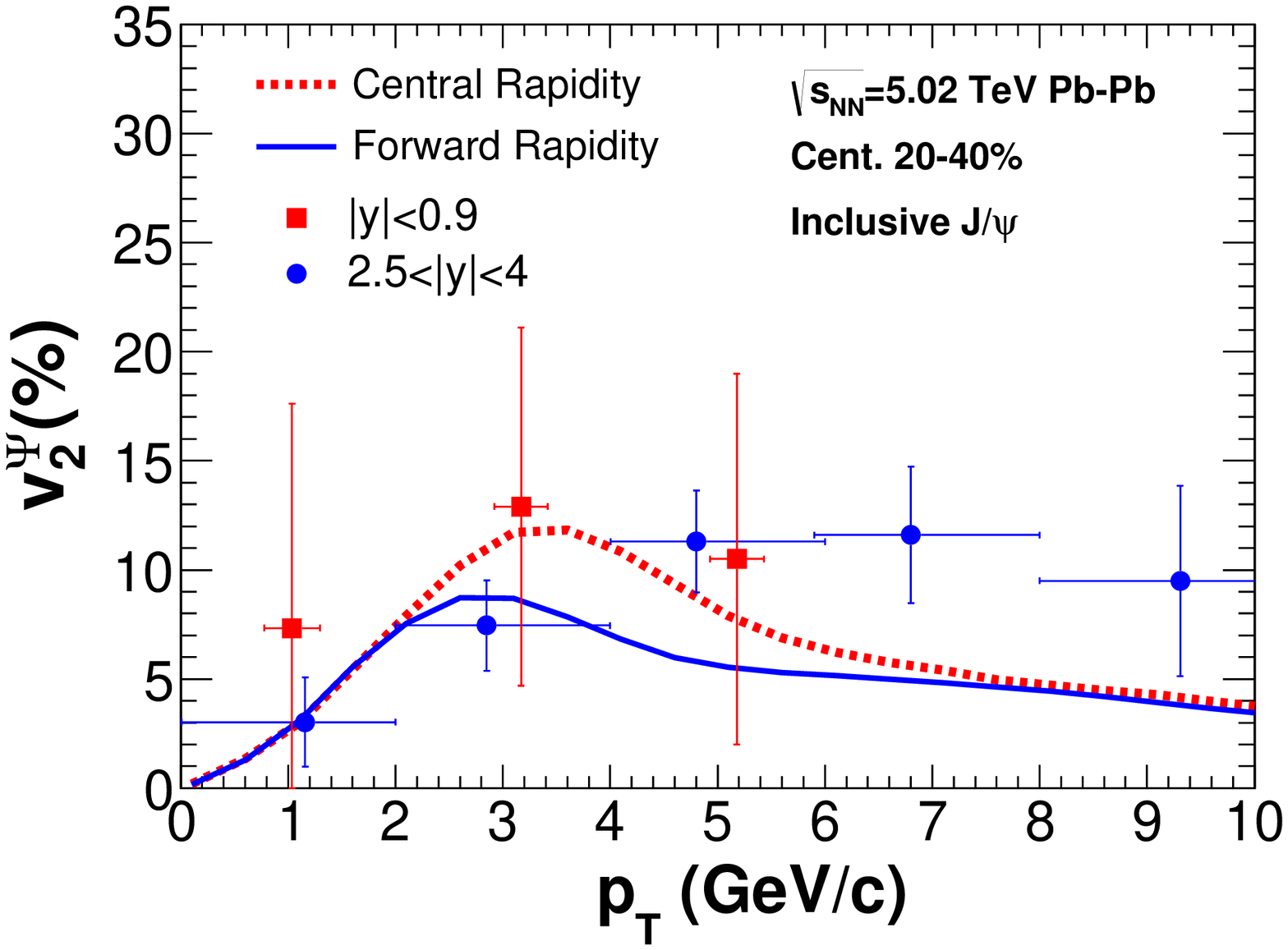}
%\includegraphics[width=0.4\textwidth]{rR-U11Tc-Gauss.eps}
%\hbox{ } \hspace{0.3cm} \hbox{ }
%\includegraphics[width=0.4\textwidth]{rR-U20Tc-Gauss.eps}
\caption{(Color online) 
(Left panel) Elliptic flows of prompt 
$J/\psi$ and $\psi(2S)$ as a function of transverse momentum $p_T$ 
in centrality 20-40\% in the forward rapidity $2.5<|y|<4$ $\sqrt{s_{NN}}=5.02$ TeV Pb-Pb collisions. 
Solid and dotted lines are for the prompt $J/\psi$ and $\psi(2S)$.  
(Right panel) 
Elliptic flows of inclusive
$J/\psi$ as a function of transverse momentum $p_T$.
Solid line and dotted line are for the inclusive $J/\psi$ in forward and central rapidities.
Experimental data is from ALICE Collaboration~\cite{Acharya:2017tgv}.
}
\hspace{-0.1mm}
\label{labv2-both}
\end{figure*}

Further more, the anisotropies of $J/\psi$ and $\psi(2S)$ momentum 
distributions are also studied in Fig.\ref{labv2-both}. 
Charmonia moving inside QGP likely as a color neutral bound state, 
are weakly coupled with the bulk medium, 
and therefore less affected by the collective expansions of QGP 
compared with open charm quarks. 
The non-zero elliptic flow of primordially 
produced $J/\psi$ 
at $p_T>6$ GeV/c is mainly 
from the effects of path length difference in the transverse plane, see the 
solid line in left panel of Fig.\ref{labv2-both}. At the low $p_T$, 
$J/\psi$ production is dominated by the regeneration. 
These heavy quarks 
are strongly coupled with the thermal medium and inherit collective flows, which results in a peak of $v_2$ 
at $p_T\sim 3$ GeV/c. The elliptic flows of prompt $\psi(2S)$ are also presented with 
dotted line. 
As $\psi(2S)$ is regenerated in the later stage of QGP anisotropic 
expansions, their elliptic flow is larger than $J/\psi$'. 
In the high $p_T$ region, the momentum anisotropy of $\psi(2S)$ is larger than $J/\psi$', 
as they are easily dissociated and sensitive to the anisotropy of the bulk medium. 

In right panel of Fig.\ref{labv2-both}, 
the experimental data is for inclusive $J/\psi$ including non-prompt contribution 
from B hadron decay. The solid line is for inclusive $J/\psi$ 
assuming kinetic 
equilibrium for bottom quarks as an up-limit~\cite{Zhou:2014kka}. 
Non-prompt part becomes important at high $p_T$ and therefore kinetically thermalized 
bottom quarks can enhance the inclusive $v_2^{J/\psi}$ by $\sim 2\%$ at $p_T\sim 8$ GeV/c. $v_2^{J/\psi}$ in 
central rapidity is also calculated with dotted line for comparison. 
For the situation of inclusive $\psi(2S)$, 
it is connected with the energy loss of bottom quarks in QGP, and 
has been elaboratively studied in Ref.\cite{Chen:2013wmr}. 

 \section{summary}

In summary, this work employs 
the improved transport model to study the thermal production of $J/\psi$ and 
$\psi(2S)$ in Pb-Pb collisions at $\sqrt{s_{NN}}=5.02$ TeV. 
Charmonium nuclear modification 
factors are dominated by the regeneration at low $p_T$ and the primordial production 
at high $p_T$ respectively. 
With different binding energies, $J/\psi$ and $\psi(2S)$ can be sequentially produced 
in the different stage of QGP anisotropic expansions. This results in different $p_T$ 
distributions of regenerated $J/\psi$ and $\psi(2S)$. We explains well both $J/\psi$ and 
$\psi(2S)$ $R_{AA}$ and their ratio at 5.02 TeV Pb-Pb collisions.   
This clearly shows how the open charm quark evolutions can affect the final 
charmonium productions. The sequential regeneration of $J/\psi$ and $\psi(2S)$ contains 
the histories of charm quark diffusions and QGP expansions in heavy-ion collisions.

\vspace{0.5cm}
{\bf Acknowledgement:} I acknowledge instructive discussions with Prof. Pengfei Zhuang and Jiaxing Zhao.  
I am also grateful to Prof. Carsten Greiner for the kind hospitality during this study. 
This work is supported by 
NSFC Grant No. 11705125 and Sino-Germany (CSC-DAAD) Postdoc Scholarship.

\end{document}